\documentclass[11pt]{article}
\usepackage{jcapmod}

\usepackage{booktabs}
\usepackage[english]{babel}
\usepackage{amsmath,amssymb,amsbsy,amstext, amsthm, simplewick}
\usepackage{hyperref}
\usepackage{graphicx}
\usepackage{amsfonts}
\usepackage{amssymb}
\usepackage{upgreek}
\usepackage{simplewick}
 \usepackage{exscale,relsize}

\usepackage[margin=1cm,labelfont={sf,bf,scriptsize},textfont={sl,scriptsize}]{caption}

\usepackage{colortbl}
\definecolor{lightgreen}{cmyk}{0.2, 0, 0.2, 0.2}
\definecolor{lightgray}{cmyk}{0.1,0.2,0,0.1}
\definecolor{lightgray2}{cmyk}{0.1,0.1,0,0.1}

\makeatletter
\newlength{\apb@width}
\newcommand{\autoparbox}[2][c]{\settowidth{\apb@width}{#2}\parbox[#1]{\apb@width}{#2}}

\makeatother

\numberwithin{equation}{section}

\def\beq{\begin{equation}}
\def\eeq{\end{equation}}

\def\bea{\begin{eqnarray}}
\def\eea{\end{eqnarray}}

\def\d{{\rm d}}

\def\bigoh{{\mathcal O}}

\newcommand\lsim{\mathrel{\rlap{\lower4pt\hbox{\hskip1pt$\sim$}}
        \raise1pt\hbox{$<$}}}
\newcommand\gsim{\mathrel{\rlap{\lower4pt\hbox{\hskip1pt$\sim$}}
        \raise1pt\hbox{$>$}}}

\def\beq{\begin{equation}}
\def\eeq{\end{equation}}
\def\bea{\begin{eqnarray}}
\def\eea{\end{eqnarray}}

\def\d{{\rm d}}

\def\d{{\rm d}}
\def\l{{\ell}}

\def\h{{\rm h}}
\def\m{{\rm m}}

\def\k{{\boldsymbol{k}}}
\def\q{{\boldsymbol{q}}}

\def\x{{\boldsymbol{x}}}

\def\M{\mathsmaller{M}}
\def\MW{\mathsmaller{MW}}
\def\N{\mathsmaller{N}}
\DeclareRobustCommand{\SkipTocEntry}[4]{}

\def\fnl{f_{\mathsmaller{\rm NL}}}
\def\gnl{g_{\mathsmaller{\rm NL}}}
\def\tnl{\tau_{\mathsmaller{ \rm NL}}}

\begin{document}

\begin{titlepage}

\setcounter{page}{1} \baselineskip=15.5pt \thispagestyle{empty}

\bigskip\

\vspace{2cm}
\begin{center}
{\fontsize{16}{28}\selectfont  \bf On the Equivalence of Barrier Crossing, \\[8pt] Peak-Background Split, and Local Biasing}
\end{center}

\vspace{0.2cm}

\begin{center}
{\fontsize{13}{30}\selectfont  Simone Ferraro$^{\diamondsuit}$, Kendrick M.~Smith$^{\diamondsuit,\heartsuit}$, Daniel Green$^{\clubsuit, \blacklozenge, \spadesuit}$, and Daniel Baumann$^{\bigstar}$}
\end{center}


\begin{center}

\vskip 8pt
\textsl{$^\diamondsuit$ Princeton University Observatory, Peyton Hall, Ivy Lane, Princeton, NJ 08544, USA}

\vskip 7pt
\textsl{$^\heartsuit$ Perimeter Institute for Theoretical Physics, Waterloo, ON N2L 2Y5, Canada}

\vskip 7pt
\textsl{$^\clubsuit$ School of Natural Sciences,
 Institute for Advanced Study,
Princeton, NJ 08540, USA}

\vskip 7pt
\textsl{$^ \blacklozenge$
Stanford Institute for Theoretical Physics, Stanford University, Stanford, CA 94306, USA}

\vskip 7pt
\textsl{$^\spadesuit$ Kavli Institute for Particle Astrophysics and Cosmology, Stanford, CA 94025, USA}

\vskip 7pt
\textsl{$^\bigstar$ D.A.M.T.P., Cambridge University, Cambridge, CB3 0WA, UK}

\end{center}

\vspace{1.2cm}
\hrule \vspace{0.3cm}
{ \noindent \textbf{Abstract} \\[0.2cm]
\noindent 
Several, apparently distinct, formalisms exist in the literature for predicting
the clustering of dark matter halos. It has been noticed on a case-by-case basis that the predictions of these
different methods agree in specific examples, but there is no general
proof that they are equivalent.  In this paper, we give a simple proof of the mathematical equivalence of barrier crossing, peak-background split, and local biasing.}  
 \vspace{0.3cm}
 \hrule

\vspace{0.6cm}
\end{titlepage}

 \tableofcontents

\newpage

\section{Introduction}

The large-scale clustering of dark matter halos has become an important probe of primordial cosmology.
In particular, non-Gaussianity in the initial conditions 
would leave an imprint in the scale-dependence of the halo bias~\cite{Dalal:2007cu, Matarrese:2008nc}, sometimes of stochastic type~\cite{Tseliakhovich:2010kf, stochastic}. Several, apparently distinct, methods are commonly used to compute these effects. 
So far, these methods have been considered to be independent, 
even though they 
give the same results when applied to specific examples~\cite{Desjacques:2011mq,stochastic}.
In this paper, we will show that the barrier crossing (BC) model, the peak-background split (PBS) method and the local biasing (LB) approach are, in fact, mathematically equivalent.

\vskip 4pt
{\it Barrier crossing}\ is the classic model of structure formation dating back to the pioneering work of Press and Schechter \cite{Press:1973iz}. In its simplest formulation, it identifies halos as regions of the linearly evolved density field above some critical density $\delta_c$. 
The clustering properties of halos can then be calculated as an 
Edgeworth expansion in the cumulants of the probability density 
of the primordial density fluctuations, 
which in turn can be expressed in terms of $N$-point functions of the 
potential~\cite{LoVerde:2007ri,Desjacques:2011mq, Smith:2011ub, stochastic}.

\vskip 4pt
{\it Peak-background split}\, is a method for calculating the influence of long-wavelength fluctuations (larger than the halo size) on the locally measured statistical properties. It has been widely used in cosmology~\cite{Bardeen:1986, Cole:1989} and its usefulness in dealing with non-Gaussian initial conditions has been first pointed out in~\cite{Dalal:2007cu}. In the most common implementation, the non-Gaussian field is defined as a non-linear function of auxiliary Gaussian fields, which are split into short-wavelength and long-wavelength components. 
By modulating the statistics of the short modes, the long modes affect the clustering statistics.
In this paper, we will generalize the PBS approach so that it can be applied to arbitrary non-Gaussian initial conditions, parametrized by arbitrary $N$-point functions of the primordial potential. 
This will require introducing additional fields $\rho_2, \rho_3, \cdots$, which measure the local power spectrum amplitude, skewness, etc. 

\vskip 4pt
{\it Local biasing}~\cite{Fry:1993, Sefusatti:2009, Giannantonio:2009, Baldauf:2011, Scoccimarro:2011pz} refers to the idea of expressing the halo density field $\delta_\h$ as a function the local dark matter density (smoothed on some scale) and expanding in powers of the density contrast~$\delta$,
\beq
\delta_{\h}(\x) = b_1 \delta(\x) + b_2 \delta^2(\x) + b_3 \delta^3(\x) + \cdots\ .
\eeq
Correlation functions can then be computed straightforwardly in terms of the coefficients in the expansion. Several variations of this formalism exist in literature (for example some use an expansion in the non-linear dark matter density, while others use the linearly evolved density). In this work, we will demonstrate the equivalence between barrier crossing and a particular variant of local biasing, in which the expansion is in the linearly evolved and non-Gaussian dark matter density contrast.

\vskip 4pt
In a companion paper~\cite{stochastic}, we derived the clustering statistics for specific non-Gaussian models,  both in the peak-background split formalism and in the barrier crossing model. 
We showed for each example that both approaches give consistent results. The goal of this paper is to prove that this agreement isn't accidental, but follows from the mathematical equivalence of both methods.

\vskip 4pt
The outline of the paper is as follows. 
After defining our notation in Section~\ref{sec:preliminaries}, we introduce our main technical tool
in Section~\ref{sec:series_representation}: a series expansion for the halo field $\delta_{\h}$ in
the barrier crossing model.
We review some examples of non-Gaussian models and show how the series expansion is used for efficiently calculating halo power spectra.
In Section~\ref{sec:equivalence}, we use the series expansion to prove the mathematical equivalence of the barrier crossing model, the peak-background split method, and the local biasing formalism. 
We conclude with brief comments in Section~\ref{sec:discussion}. An appendix collects some elementary properties of Hermite polynomials.

\section{Preliminaries and Notation}
\label{sec:preliminaries}


Non-Gaussian initial conditions can be parameterized by the connected $N$-point functions $\xi_\Phi^{(N)}$ of the primordial
gravitational potential $\Phi$.  In Fourier space, these are defined as
\beq
\langle \Phi_{{\k}_1}  \Phi_{{\k}_2} \cdots  \Phi_{{\k}_N}\rangle_{\rm c} 
  = (2\pi)^3 \delta_{\rm D}({\k}_{1 2\dots N}) \, \xi_\Phi^{(N)}({\k}_1, {\k}_2, \dots, {\k}_N)\ , \label{equ:xi}
\eeq
where ${\k}_{1 2\dots N} \equiv {\k}_{1} + {\k}_{2} + \cdots + {\k}_{N}$.
The primordial potential is related to the linearly evolved matter density contrast via Poisson's equation
\beq
\delta_\k(z) = \alpha(k,z) \Phi_\k\ , \label{equ:dk}
\eeq 
where 
\beq
\alpha(k,z) \equiv \frac{2 \hskip 1pt k^2 T(k) D(z)}{3\hskip 1pt  \Omega_m H_0^2} \ .
\eeq
Here, $T(k)$ is the matter transfer function normalized such that $T(k) \rightarrow 1$ as $k \rightarrow 0$ and $D(z)$ is the linear growth factor (as function of redshift $z$), normalized so that $D(z) = (1+z)^{-1}$ in matter domination.  For notational simplicity, we will from now on suppress the redshift argument from all quantities.
The field $\delta_\M(\x)$ denotes the linear density contrast smoothed with a top-hat filter of radius $R_\M = (3 M/ 4 \pi \bar{\rho}_{\rm{m}})^{1/3}$. 
In Fourier space, 
\beq
\delta_\M(\k) = W_\M(k) \delta_\k\ ,
\eeq 
where $W_\M(k)$ is the Fourier transform of the top-hat filter,
\beq
W_\M(k) \equiv 3\, \frac{\sin(kR_\M) - kR_\M \cos(kR_\M)}{(kR_\M)^3}\ .
\eeq
We also define $\sigma_\M \equiv \langle \delta_{\M}^2 \rangle^{1/2}$ and $\alpha_\M(k) \equiv W_\M(k) \alpha(k)$.

\vskip 4pt
 The main quantity of interest, in this paper, is the halo density contrast in Lagrangian space
 \beq
 \delta_{\h}(\x) \equiv \frac{n_\h(\x) - \langle n_\h \rangle}{\langle n_\h \rangle}\ ,
 \eeq 
 where $n_\h(\x)$ is the halo number density.
To lowest order, $\delta_\h$ is related to the halo overdensity in Eulerian space via $\delta_{\h}^{\mathsmaller{\rm E}} = \delta_{\h} + \delta$.
We will determine the large-scale behavior of the matter-halo and halo-halo power
spectra $P_{\m\h}(k) \equiv \langle \delta \delta_\h \rangle(k)$ and $P_{\h\h}(k) \equiv \langle \delta_\h \delta_\h \rangle(k)$.
We define $P_{\h\h}(k)$ to be the halo power spectrum after the
shot noise contribution $1/n_\h$ has been subtracted, where $n_\h$ is the halo number density.
Analogously, we define $P_{\m\h}(k)$ to be the matter-halo power spectrum after subtracting the 1-halo term (in practice, this
term is usually negligibly small).
We define the (Lagrangian) halo bias as
\beq
b(k) \equiv \frac{P_{\m\h}(k)}{P_{\m\m}(k)}\ .
\eeq 
This is related to the Eulerian bias via $b_{\mathsmaller{\rm E}} = b+1$.
A stochastic form of halo bias arises whenever the density of halos isn't 100\% correlated with the dark matter density~\cite{stochastic}.
In that case, the bias inferred from $P_{\h\h}$ will be different from the bias inferred from $P_{\m\h}$, i.e.
\beq
 \frac{P_{\h\h}(k)}{P_{\m\m}(k)} \ \ne\ \left(  \frac{P_{\m\h}(k)}{P_{\m\m}(k)} \right)^2\ .
\eeq

\section{A Series Representation of Barrier Crossing}
\label{sec:series_representation}


In this section, we introduce the barrier crossing formalism and quote results from our companion paper~\cite{stochastic}.  We also introduce a series representation of barrier crossing, which will be our main tool to prove the equivalence to local biasing and peak-background split in Section~\ref{sec:equivalence}.

\subsection{Review of Barrier Crossing} 
\label{ssec:barrier_crossing_model}

In the simplest version of the barrier crossing model~\cite{Press:1973iz}, 
halos of mass $\geq M$ are modeled as regions of space in which the smoothed density field~$\delta_\M$ 
exceeds the collapse threshold $\delta_c \approx 1.4$, i.e.~the halo number density $n_\h(\x)$ is
given by
\beq
n_\h^{\MW}({\x}) \propto \Theta(\delta_\M({\x}) - \delta_c)\ ,   \label{eq:nh_barrier}
\eeq
where $\Theta$ is the Heaviside step function.
Eq.~(\ref{eq:nh_barrier}) models the abundance of a {\it mass-weighted} sample of halos whose mass exceeds some minimum
value $M$.\footnote{This type of sample is often assumed when fitting models to observations of luminous tracers
such as galaxies or quasars.  In the absence of detailed knowledge of the halo occupation distribution (HOD), a simple
choice is to assume that halos below some minimum mass $M$ are unpopulated with tracers, whereas the expected number of
tracers in a halo of mass $\ge M$ is proportional to the halo mass.}  We will also consider the case of
a halo sample defined by a narrow mass bin, which is obtained from the mass-weighted case by differentiating with respect to
$M$, i.e.
\beq
n_{\h}^\N({\x}) \propto \frac{\partial}{\partial M} \Theta(\delta_{\M}({\x}) - \delta_c)\ .
\eeq
Throughout the paper, we will refer to these two types of halo samples as ``mass-weighted samples''~($MW$) and ``narrow samples''~($N$).

The barrier crossing model allows us to compute the statistics of halo-halo and halo-matter correlations.
To discuss correlations between quantities at two points $\x$ and $\x'$, it is useful to define $\delta_\M = \delta_\M(\x), \delta'_\M = \delta_\M(\x')$ and $r = | \x - \x'|$.
The joint cumulants of the density fields are then\footnote{Note that the variance of the unsmoothed linear density contrast $\sigma^2 = \langle \delta^2 \rangle$ is formally infinite, 
but cancels in the definition~(\ref{eq:f1n_def}) of the quantity $f_{\hat 1,n}$ which will appear in our final expressions.} 
\begin{align}
\kappa_{\hat m,n}(r,M) &\ \equiv\ \frac{\langle  \delta^m (\delta_{\M}')^n\rangle_{\rm c}}{ \sigma^m\sigma_{\M}^n}\ , \\
\kappa_{m,n}(r,M,\bar M) &\ \equiv\ \frac{\langle (\delta_{\M})^m (\delta_{\bar \M}')^n\rangle_{\rm c}}{\sigma_{\M}^{m} \sigma_{\bar \M}^n}  \ . 
\end{align}
The hat on $\kappa_{\hat m, n}$ denotes the use of the unsmoothed density field $\delta$.
In the limit $k \to 0$, we find $\kappa_{\hat 1,1}(k) \to P_{\m \m}(k)/(\sigma \sigma_\M)$ and $\kappa_{1,1}(k) \to P_{\m \m}(k)/(\sigma_\M \sigma_{\bar \M})$.
This motivates the following definitions
\bea
f_{\hat 1, n}(k,M) &\equiv& \frac{\kappa_{\hat 1, n}(k,M)}{\kappa_{\hat 1, 1}(k,M) \, \sigma_\M} \ \, \hskip 1pt \qquad \quad \ \mbox{for $n \ge 1$}\ , \label{eq:f1n_def} \\ 
f_{1,n}(k,M,\bar M) &\equiv& \frac{\kappa_{1,n}(k,M,\bar M)}{\kappa_{1,1}(k,M,\bar M) \sigma_{\bar\M}} \ \, \hskip 2pt \qquad  \mbox{for $n \ge 1$}\ ,  \\
f_{m,n}(k,M,\bar M)  &\equiv& \frac{\kappa_{m,n}(k,M,\bar M)}{\kappa_{1,1}(k,M,\bar M) \sigma_{\M} \sigma_{\bar \M}} \quad \   \mbox{for $m,n \ge 2$}\ . \label{eq:fmn_def} 
\eea
Using the function $\alpha(k,z)$ defined in (\ref{equ:dk}),
it is straightforward to relate the above cumulants to the primordial correlation functions $\xi_\Phi^{(N)}$ defined in (\ref{equ:xi}).

In~\cite{stochastic}, we showed how the matter-halo and halo-halo power spectra are computed in the barrier crossing model using the Edgeworth expansion for the joint probability density function~$p(\delta_{\M}, \delta_{\M}')$. (We refer the reader to that paper for detailed derivations and further discussion.)
The result can be expressed in terms of the cumulants $f_{\hat 1,n}$ and $f_{m,n}$.  
Taking the limit $k \to 0$ for the case of a mass-weighted sample with $M = \bar M$, we find
\begin{subequations}
\label{eq:edgeworth_mw0}
\begin{align}
P_{\m\h}(k,M) &= P_{\m\m}(k) \Bigg( b_g^{\MW}(M) + \sum_{n\ge 2} \alpha_n(M) f_{\hat 1,n}(k,M) \Bigg) \ ,  \\
P_{\h\h}(k,M) &= P_{\m\m}(k) \Bigg( b_g^{\MW}(M)^2
   +\, 2\hskip 1pt b_g^{\MW}(M) \sum_{n \ge 2}  \alpha_n(M) f_{1,n}(k,M,M)\label{eq:edgeworth_mw}   \\
  & \hspace{3.7cm} + \sum_{m,n\ge 2} \alpha_m(M) \alpha_n(M) f_{m,n}(k,M,M) \Bigg)\ ,  \nonumber
\end{align}
\end{subequations}
where the coefficients $\alpha_n$ (not to be confused with the $\alpha$ of eq.~(\ref{equ:dk})) are defined in terms of Hermite polynomials (see Appendix~\ref{sec:hermite}),
\beq
\alpha_n(M) \equiv \sqrt{\frac{2}{\pi}} \, \frac{e^{-\nu_c^2/2}}{{\rm erfc}(\frac{1}{\sqrt{2}}\nu_c)} \frac{H_{n-1}(\nu_c)}{n!}\ , \quad {\rm with} \quad  \nu_c(M) \equiv \frac{\delta_c}{\sigma_{\M}}\ . \label{eq:alphan_def}
\eeq
We also defined the Gaussian bias as
\beq
b_g^{\MW}(M) \equiv \frac{\alpha_1(M)}{\sigma_{\M}}  \ . \label{eq:bgMW_def}
\eeq
Note that $b_g^{\MW}(M)$ is the Press-Schechter bias for
the mass-weighted halo sample.  In writing~(\ref{eq:edgeworth_mw0}), we have dropped  ``nonlinear'' terms in the Edgeworth expansion, i.e.~terms involving products 
$(\kappa_{m_1n_1} \kappa_{m_2n_2} \cdots \kappa_{m_pn_p})$ with $p > 1$.  

Similarly, for the case of a halo sample defined by a narrow mass bin, we have
\begin{subequations}
\begin{align}
P_{\m\h}(k,M) 
   &= P_{\m\m}(k) \left( b_g^{\N}(M) + \sum_{n \ge 2} 
        {\cal D}_n(M)  f_{\hat 1,n}(k,M) \right)\ , \\
P_{\h\h}(k,M)
   &= P_{\m\m}(k) \Bigg( b_g^{\N}(M)^2 +\, 2\hskip 1pt b_g^{\N}(M) \sum_{n \ge 2} 
       {\cal D}_n(M)   f_{1,n}(k,M,\bar M) \Big|_{M=\bar M}  \label{eq:phh_edgeworth_narrow} \\
    & \hspace{3.4cm} + \sum_{m,n\ge 2} 
          {\cal D}_m(M)  
          {\cal D}_n(\bar M)  
          f_{m,n}(k,M,\bar M) \Big|_{M=\bar M} \Bigg)\ ,  \nonumber 
\end{align}
\end{subequations}
where we have defined the differential operator 
\beq
 {\cal D}_n(M) \equiv \beta_n(M) + \tilde\beta_n(M) \frac{\partial}{\partial \ln\sigma_\M}\ ,
\eeq
as well as the functions
\beq
b_g^{\N}(M) \equiv  \frac{1}{\sigma_\M}\frac{\nu_c^2 -1}{\nu_c}\quad , \quad \beta_{n}(M) \equiv \frac{H_n(\nu_c)}{n!} \quad {\rm and} \quad \tilde \beta_{n}(M) \equiv \frac{H_{n-1}(\nu_c)}{n!\,\nu_c}\ .  \label{eq:beta_def}
\eeq
Note that $b_g^{\N}(M)$ is the Press-Schechter bias of a halo sample defined by a narrow mass bin.
In eq.~(\ref{eq:phh_edgeworth_narrow}) for $P_{\h\h}$, we have assumed $M=\bar M$ for simplicity, but the variables $M$ and $\bar M$ should be treated
as independent for purposes of taking derivatives.

\subsection{Hermite Polynomial Expansion}

In this section, we will develop an alternative (to the Edgeworth expansion) algebraic
framework for analyzing clustering in the barrier crossing model.
First, consider the case of a mass-weighted halo sample, where the halo field is modeled as a step function
\beq
n_\h^\MW(\x) \propto \Theta\left( \nu(\x) - \nu_c \right)\ , \quad {\rm where} \quad \nu(\x) \equiv \frac{\delta_{\M}(\x)}{\sigma_{\M}}\ .  \label{eq:herm1}
\eeq
Since the Hermite polynomials $H_n(\nu)$ are a complete basis, any function of $\nu$ can be
written as a linear combination of Hermite polynomials.  In particular, we can write the Heaviside
step function $\Theta(\nu-\nu_c)$ as
\beq
\Theta(\nu-\nu_c) = \sum_{n=0}^\infty a_n(\nu_c) \, H_n(\nu) \ ,  \label{eq:herm2}
\eeq
where
\beq
a_n(\nu_c) = \frac{1}{n!} \int_{-\infty}^\infty \d\nu\ \Theta(\nu-\nu_c)\, \frac{e^{-\nu^2/2}}{\sqrt{2\pi}} \, H_n(\nu) 
  = \left\{ \begin{array}{ll} 
        \tfrac{1}{2}{\rm erfc}(\tfrac{1}{\sqrt{2}} \nu_c) & \quad n=0 \\ 
        \tfrac{1}{n!} \frac{1}{\sqrt{2\pi}} e^{-\nu_c^2/2}\, H_{n-1}(\nu_c)& \quad n\ge 1
    \end{array} \right. \ .  \label{eq:explicit_an}
\eeq
Plugging this series expansion into eq.~(\ref{eq:herm1}), and normalizing the halo field to the fractional overdensity $\delta_{\h}$, we get
\begin{eqnarray}
\delta_{\h}(\x) &=& \sum_{n \ge 1} \frac{a_n(\nu_c)}{a_0(\nu_c)} H_n\left( \frac{\delta_{\M}(\x)}{\sigma_{\M}} \right) \nonumber \\
   &=& b_g^{\MW}(M) \delta_{\M}(\x) + \sum_{n\ge 2} \alpha_n(M) \rho_n(\x) \ , \hspace{1cm} \mbox{[mass-weighted sample]}  \label{eq:series_mw}
\end{eqnarray}
where 
$\alpha_n(M)$ and $b_g^{\MW}(M)$ were introduced in eqs.~(\ref{eq:alphan_def}) and (\ref{eq:bgMW_def}), respectively.
The fields $\rho_n$ are defined as
\beq
\rho_n({\x}) \equiv H_n\!\left(\frac{\delta_\M({\x})}{\sigma_\M}\right)\ .
\eeq
On large scales, the field $\rho_2 = \delta_{\M}^2/\sigma_{\M}^2 - 1$ tracks long-wavelength variations in the locally measured small-scale power,
and for non-Gaussian initial conditions the power spectrum $P_{\rho_2 \rho_2}(k)$ may acquire extra large-scale contributions.
Analogously, the field $\rho_3 = \delta_{\M}^3/\sigma_{\M}^3 - 3 \delta_\M/\sigma_{\M}$ tracks long-wavelength variations in the locally measured small-scale
skewness, and so on for higher $\rho_n$.

\begin{figure}[h!]
   \centering
    \hspace{-2cm}   \includegraphics[width=10.5cm]{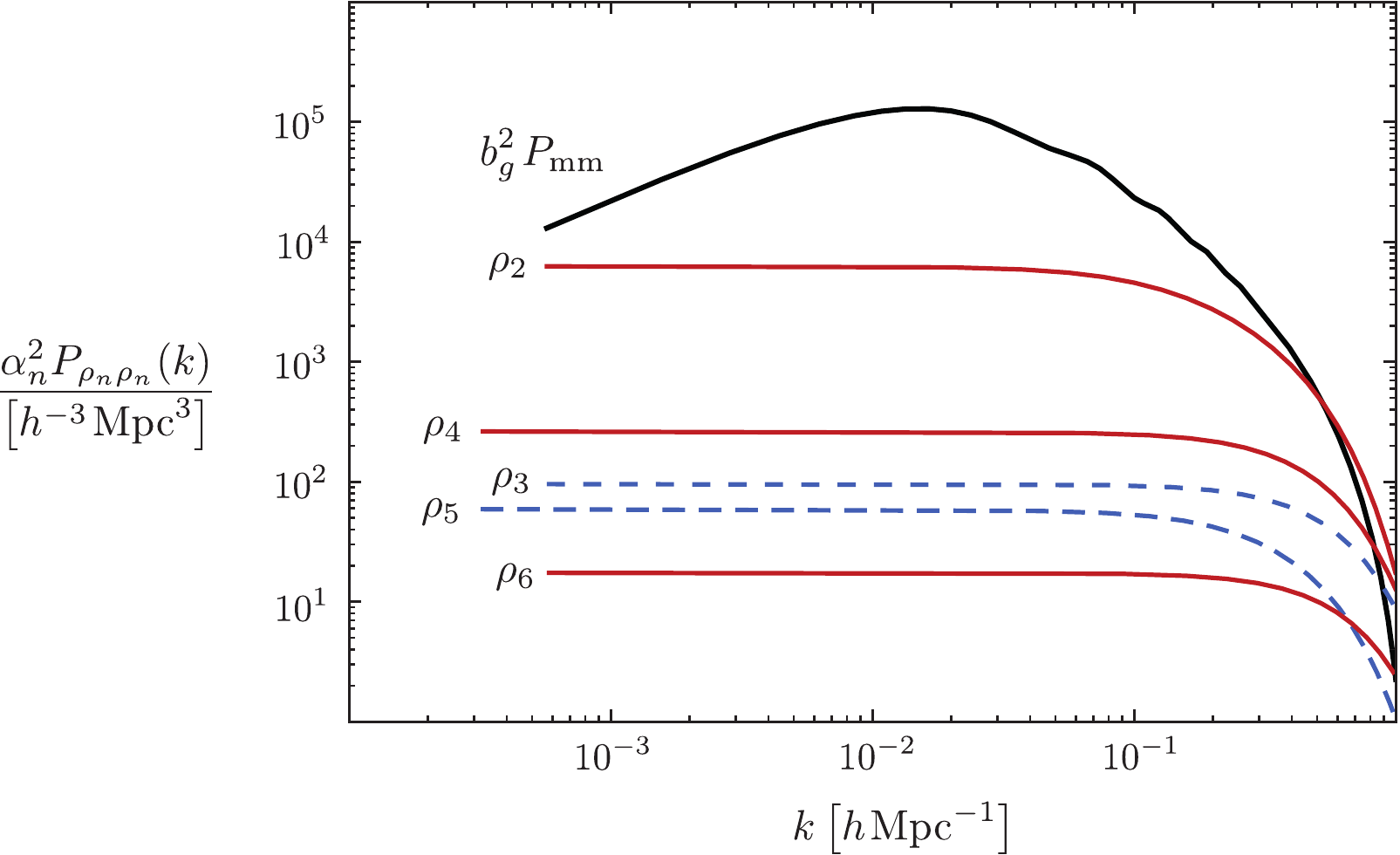}
   \caption{Convergence of the series representation~(\ref{eq:series_mw}) at low $k$, illustrated by comparing
terms in the halo-halo power spectrum $P_{\h\h}(k) = b_g^2(M) P_{\m \m}(k) + \sum_{n=2}^\infty \alpha_n^2(M) P_{\rho_n\rho_n}(k)$
in a Gaussian cosmology.  (Note that for Gaussian initial conditions, cross power spectra $P_{\rho_m\rho_n}(k)$ with $m\ne n$
are zero.)  We have taken $z=0$ and a mass-weighted sample of halos with mass $M \ge 2\times 10^{13}$ $h^{-1}$ $M_\odot$.}
  \label{fig:rho}
\end{figure}

We emphasize that the series representation~(\ref{eq:series_mw}) is mathematically equivalent to the barrier crossing model,
since it is obtained by simply substituting the convergent Hermite series~(\ref{eq:herm2}) into the barrier crossing expression~(\ref{eq:herm1})
for $n_\h$.
The series representation converges for all values of $\x$, but
its usefulness depends on how rapidly it converges, i.e.~how many terms we need to get a good approximation.
For example, to compute the halo field $\delta_\h(\x)$ at a single point $\x$ in real space, many terms are needed (of order 100) and the series
representation is not useful.
On the other hand, the Fourier transformed series representation $\delta_\h(\k) = b_g^{\MW} \delta_\M +\sum_{n=2}^\infty \alpha_n \rho_n(\k)$
converges rapidly on large scales (i.e.~$k \ll k_{\rm nl}$), as shown in fig.~\ref{fig:rho}, and the series representation is very convenient.
(The series converges for all $k$, but only converges rapidly for
$k \ll k_{\rm nl}$.)  

The preceding expressions have all applied to the case of a mass-weighted halo sample.
For the case of a halo sample defined by a narrow mass bin, the halo field is modeled as
\begin{eqnarray}
n_\h^\N(\x) 
  & \propto & \frac{\partial}{\partial \ln \sigma_\M} \Theta\left( \frac{\delta_{\M}(\x)}{\sigma_{\M}} - \nu_c \right) \nonumber \\
  & = & \frac{\partial}{\partial\ln\sigma_\M} \sum_{n\ge 0} a_n(\nu_c) H_n\!\left( \frac{\delta_\M(\x)}{\sigma_\M} \right) \nonumber \\
  & = & \sum_{n\ge 0} \left( (n+1) \nu_c \, a_{n+1}(\nu_c) + a_n(\nu_c) \frac{\partial}{\partial\ln\sigma_\M} \right) 
           H_n\!\left( \frac{\delta_\M(\x)}{\sigma_\M} \right) \ .
\end{eqnarray}
Normalizing $n_{\h}$ to the fractional halo overdensity $\delta_{\h}$, we get
\begin{eqnarray}
\delta_\h(\x) & = &
  \sum_{n\ge 1} \left( (n+1) \frac{a_{n+1}(\nu_c)}{a_1(\nu_c)}
       + \frac{a_n(\nu_c)}{\nu_c \hskip 1pt a_1(\nu_c)} \frac{\partial}{\partial\ln\sigma_\M} \right) H_n\!\left( \frac{\delta_\M(\x)}{\sigma_\M} \right) \nonumber \\
  &=& \left( \frac{\nu_c^2 - 1}{\nu_c\hskip1pt \sigma_\M } \right) \delta_\M(x) + \frac{1}{\nu_c \hskip 1pt \sigma_\M} \frac{\partial \delta_\M(\x)}{\partial\ln\sigma_\M} \nonumber \\
   && \hspace{2.8cm}
       + \sum_{n\ge 2} \left( \frac{1}{n!} H_n(\nu_c) + \frac{1}{n!} \frac{H_{n-1}(\nu_c)}{\nu_c} \frac{\partial}{\partial\ln\sigma_\M} \right)
           H_n\!\left( \frac{\delta_\M(\x)}{\sigma_\M} \right) \ . \ \ \ 
\end{eqnarray}
We drop the term containing $\partial\delta_{\M}/\partial\ln\sigma_{\M}$, since this term vanishes on large scales, $k \ll R_{\M}^{-1}$,
and write the result using the notation $b_g, \beta_n, \tilde\beta_n$ defined in eq.~(\ref{eq:beta_def}):
\beq
\delta_{\h}(\x) = b_g^{\N}(M) \delta_{\M}(\x) + \sum_{n=2}^\infty \left( \beta_n(M) + \tilde\beta_n(M) \frac{\partial}{\partial\ln \sigma_{\M}} \right) \rho_n(\x)\ . \label{eq:series_narrow}
  \hspace{1cm} \mbox{[narrow sample]}
\eeq
As a check on our formalism, we can verify that the matter-halo and halo-halo power spectra obtained from the series~(\ref{eq:series_mw})
agree with the results obtained previously in~\cite{stochastic} using the Edgeworth expansion.
We first write the power spectrum $P_{\delta\rho_n}(k)$ in terms of the correlation function~(\ref{equ:xi}), 
\begin{eqnarray}
P_{\delta\rho_n}(k) 
  &=& \frac{\alpha(k)}{\sigma_{\M}^n} \left( \prod_{i=1}^{n-1} \int_{\q_i} \alpha_{\M}(q_i) \right) \alpha_\M(-|\k+\q|)\, \times \ \xi^{(n+1)}_\Phi(\k,\q_1, \cdots,\q_{n-1}, - \k - \q)  \ ,  \ \  \ \  \label{equ:P1}
       \end{eqnarray}
 where we have defined $\int_{\q_i}\, (\cdot) \equiv \int \frac{\d^3 \q_i}{(2\pi)^3}\, (\cdot)$ and $\q \equiv \sum_{i=1}^{n-1} \q_{i} $.  
 Similarly, we can express $P_{\rho_m\rho_n}(k)$ as\hskip 2pt\footnote{We have made an approximation here: 
by using {\em connected} correlation functions in eqs.~(\ref{equ:P1}) and~(\ref{equ:P2}), we have neglected some contributions to the power spectra
$P_{\delta\rho_n}$ and $P_{\rho_m\rho_n}$.  More precisely, we have neglected disconnected terms whose factorization contains multiple higher
cumulants (i.e.~$\kappa_{m,n}$ with $m+n\ge 3$), and also some contributions to $P_{\rho_m\rho_n}(k)$ which approach a constant as $k\rightarrow 0$.
(Note that subleading terms in the Hermite polynomial
$\rho_n = (\delta_{\M}/ \sigma_{\M})^n - n(n-1)(\delta_{\M}/\sigma_{\M})^{n-2}/2 + \cdots$
cancel the largest disconnected contributions to the power spectra in eqs.~(\ref{equ:P1}) and~(\ref{equ:P2}).)
The derivation in~\cite{stochastic} of eq.~(\ref{eq:edgeworth_mw0}) contains equivalent approximations, which is why
we will shortly find agreement with the results of~\cite{stochastic}.
In principle, one can avoid making any approximations by including disconnected contributions when calculating power spectra
$P_{\delta\rho_n}$ and $P_{\rho_m\rho_n}$.
However, in Appendix~A of~\cite{stochastic}, we showed that these approximations are always valid in the observationally
relevant regime where the initial perturbations are close to Gaussian.}
       \begin{eqnarray}
\hspace{-2.5cm}P_{\rho_m\rho_n}(k)
  &=& \frac{1}{\sigma_{\M}^m \sigma_{\bar {\M}}^n} 
        \left( \prod_{i=1}^{m-1} \int_{\q_i'} \alpha_\M(q_i') \right) 
        \left( \prod_{j=1}^{n-1} \int_{\q_j} \alpha_{\bar {\M}}(q_j) \right) \alpha_\M(q') \alpha_{\bar {\M}}(q) \nonumber \\
   && \hspace{2cm}
       \times \ \xi_\Phi^{(m+n)}(\q_1',\cdots,\q_{m-1}',-\q'+\k,\q_1,\cdots,\q_{n-1},-\q-\k) \ , \label{equ:P2}
       \end{eqnarray}
       where $\q' \equiv \sum_{i=1}^{m-1} \q_{i}' $\hskip 1pt.
 Using the notation
$f_{\hat 1,n}$ and $f_{m,n}$ defined in eqs.~(\ref{eq:f1n_def}) and (\ref{eq:fmn_def}), and taking the limit $k \to 0$, we find
\begin{eqnarray}
P_{\delta\rho_n}(k) 
  &=& f_{\hat 1,n}(k,M) \, P_{\m\m}(k) \ ,\\
P_{\rho_m\rho_n}(k)  &=& f_{m,n}(k,M,\bar M) \, P_{\m\m}(k)\ .
\end{eqnarray}
For the case of the mass-weighted halo sample, the series representation~(\ref{eq:series_mw}) therefore gives the following matter-halo and halo-halo power spectra
\begin{subequations}
\begin{eqnarray}
P_{\m\h}(k,M) &=& P_{\m\m}(k) \left( \sum_{n\ge 1} \alpha_n(M) f_{\hat 1,n}(k,M) \right) \ , \\
P_{\h\h}(k,M,\bar M) &=& P_{\m\m}(k) \left( \sum_{m,n\ge 1} \alpha_m(M) \alpha_n(\bar M) f_{m,n}(k,M,\bar M) \right)\ ,
\end{eqnarray}
\end{subequations}
in agreement with the Edgeworth calculation~(\ref{eq:edgeworth_mw0}).
The case of the narrow mass bin can be verified similarly.

\vskip 4pt
Eqs.~(\ref{eq:series_mw}) and~(\ref{eq:series_narrow}) are the main results of this section and give a series representation for the halo
field in the barrier crossing model, for the cases of a mass-weighted halo sample and a narrow mass bin respectively.
Using the series representation, we will give a simple, conceptual proof of the equivalence of the barrier model, the
peak-background split, and local biasing in~Section~\ref{sec:equivalence}.
However, it is useful to first build intuition by considering a few example non-Gaussian models.

\subsection{Examples}
\label{sec:examples}

For a given non-Gaussian model, one can analyze large-scale clustering by keeping a small set of terms in the series
expansion of $\delta_{\h}$ (either eq.~(\ref{eq:series_mw}) or~(\ref{eq:series_narrow}) for a mass-weighted sample or narrow mass bin, respectively),
and computing the necessary power spectra $P_{\rho_m\rho_n}(k)$ on large scales.
This is a computationally convenient way to compute the non-Gaussian clustering signal, and allows the signal to be interpreted
physically as arising from large-scale variations in locally measured quantities such as small-scale power and skewness,
as we will see in the context of some example models.

\subsubsection{$\tau_{\rm NL}$ Cosmology}
\label{ssec:tnl}

Consider a non-Gaussian model in which the initial Newtonian potential is given by
\beq
\Phi(\x) = \phi(\x) + \fnl \left(\phi^2(\x) - \langle \phi^2 \rangle \right)\ ,
\eeq where $\phi$ is a Gaussian field.
We will refer to this as the ``\hskip 1pt $\fnl$ model\hskip 1pt" (or local model).  
This type of non-Gaussianity arises somewhat generically in multi-field models of the early universe,
e.g.~modulated reheating models~\cite{Zaldarriaga:2003my},
curvaton models~\cite{Linde:1996gt,Lyth:2001nq},
or multi-field ekpyrotic scenarios~\cite{Lehners:2007ac,Buchbinder:2007ad}.
In this section, we will study a generalization of the $\fnl$ model which we will call
the ``\hskip 1pt $\tnl$ model\hskip 1pt". This type of non-Gaussianity arises in ``multi-source'' models,
i.e.~models in which quantum mechanical perturbations in multiple fields determine the initial adiabatic
curvature perturbation~\cite{Tseliakhovich:2010kf, Chen:2009zp, Baumann:2011nk, Quasi4}.
The non-Gaussian potential $\Phi$ is given in terms of two uncorrelated Gaussian fields $\phi$ and $\psi$, 
with power spectra that are proportional to each other
\beq
\Phi(\x) = \phi(\x) + \psi(\x) + \fnl(1+\Pi)^2 \, \left(\psi^2(\x) - \langle \psi^2 \rangle \right)\ ,  \label{eqn:taunlmodel}
\eeq
where $\fnl$ and $\Pi = P_\phi(k) / P_\psi(k)$ are free parameters.
It is easy to compute the three- and four-point functions,
\bea \xi^{(3)}_{\Phi}
&=& \fnl \big[P_1 P_2 + \mathrm{5 \, perms.} \big] + \mathcal{O}(\fnl^3)\ , \label{equ:tnl3}\\
\xi^{(4)}_{\Phi}
&=& 2\left (\tfrac{5}{6} \right )^2 \tnl \big[P_1P_2 P_{13} + \mathrm{23 \, perms.} \big] + \mathcal{O}(\tnl^2) \ , \label{equ:tnl4}
\eea
where we have defined $\tnl = (\frac{6}{5} \fnl)^2 (1+\Pi)$, $P_i \equiv P_\Phi(k_i)$, and $P_{ij} \equiv P_\Phi(|{\k}_i + {\k}_j|)$.
It is conventional to parametrize this model with variables $\{ \fnl, \tnl \}$, which correspond to the amplitudes of
the 3-point and 4-point functions, rather than the variables $\{ \fnl, \Pi \}$.
The $\fnl$ model (with $\Pi=0$ so that $\psi$ contributes but not $\phi$) corresponds to the special case $\tnl = ( \frac{6}{5} \fnl )^2$.

To compute halo clustering in the $\tnl$ model, we keep the first two terms in the series 
expansion for $\delta_{\h}$ (eqs.~(\ref{eq:series_mw}) and~(\ref{eq:series_narrow})), obtaining:
\beq
\delta_{\h} = \left\{ \begin{array}{lr}
  b_g^{\MW} \delta_{\M} + \alpha_2 \rho_2 \ ,
      & \hspace{1cm} \mbox{[mass-weighted sample]}  \\[0.2cm]
 b_g^{\N} \delta_{\M} + \left( \beta_2(M) + \tilde\beta_2(M) \frac{\partial}{\partial\ln\sigma_{\M}} \right) \rho_2\ .
      & \mbox{[narrow sample]}
\end{array} \right.  \label{eq:tnl_series}
\eeq
Using eqs.~(\ref{equ:tnl3}) and (\ref{equ:tnl4}) in eqs.~(\ref{equ:P1}) and (\ref{equ:P2}), 
we obtain the following power spectra in the $k\rightarrow 0$ limit:
\begin{eqnarray}
P_{\delta\rho_2}(k) 
  &=& 4\fnl \frac{P_{\m\m}(k)}{\alpha(k)}  \ , \label{eq:tnl_rho11} \\
P_{\rho_2\rho_2}(k)
  &=& 16 \left( \tfrac{5}{6} \right)^2 \tnl \frac{P_{\m\m}(k)}{\alpha^2(k)} \ .  \label{eq:tnl_rho22}
\end{eqnarray}
Putting everything together, we find
\begin{subequations}
\label{eq:tnl_ps}
\begin{eqnarray}
P_{\m\h}(k) &=& \left( b_g + \fnl \frac{\beta_f}{\alpha(k)} \right) P_{\m\m}(k)\ ,  \\
P_{\h\h}(k) &=& \left( b_g^2 + 2 b_g \fnl \frac{\beta_f}{\alpha(k)} 
  + \left( \tfrac{5}{6} \right)^2 \tnl \frac{\beta_f^2}{\alpha^2(k)} \right) P_{\m\m}(k)\ . 
\end{eqnarray}
\end{subequations}
where we have defined the non-Gaussian bias parameter
\beq
\beta_f = \left\{ \begin{array}{lr}
   4 \alpha_2(M)\ , & \hspace{1cm} \mbox{[mass-weighted sample]} \\
   4 \beta_2(M) \ . & \hspace{1cm} \mbox{[narrow sample]} \\
\end{array} \right.  \label{eq:betaf_def}
\eeq
In both the mass-weighted and narrow mass bin cases,
the non-Gaussian and Gaussian parts of the bias are related by $\beta_f = 2 \delta_c b_g$.
Note that in the narrow mass bin case, there is a derivative term in $\delta_{\h}$
(the term $\partial\rho_2/\partial\ln\sigma_{\M}$ in eq.~(\ref{eq:tnl_series})), but this ends up giving zero
contribution to the power spectra $P_{\m\h}$ and $P_{\h\h}$, 
since the power spectra $P_{\delta\rho_2}$ and $P_{\rho_2\rho_2}$ are independent of $M$ in
the $\tnl$ model.

Our calculation of the clustering power spectra~(\ref{eq:tnl_ps}) agrees with previous 
calculations in the literature (e.g.~\cite{Tseliakhovich:2010kf,stochastic})
but the series representation gives some physical intuition: the large-scale non-Gaussian clustering is due
to large-scale fluctuations in the field $\rho_2$, which we interpret as long-wavelength variations in the locally measured
small-scale power.
If $\tnl = (\frac{6}{5} \fnl)^2$, then long-wavelength variations in $\rho_2$ are 100\% correlated to the matter density $\delta$ on large scales, 
and the non-Gaussian halo bias is non-stochastic.
If $\tnl > (\frac{6}{5} \fnl)^2$, then $\rho_2$ and $\delta$ are not 100\% correlated, leading to stochastic bias.

\subsubsection{$g_{\rm NL}$ Cosmology}
\label{ssec:gnl}

The $\gnl$ model is a non-Gaussian model in which the initial potential $\Phi$ is given
in terms of a single Gaussian field~$\phi$ by:
\beq
\Phi(\x) = \phi(\x) + \gnl \left(\phi^3(\x) - 3 \langle \phi^2 \rangle \phi(\x) \right)\  . \label{equ:gnl_model}
\eeq
We keep the first three terms in the series expansion for $\delta_{\h}$, obtaining:
\beq
\delta_{\h} = \left\{ \begin{array}{lr}
 b_g^{\MW} \delta_{\M} + \alpha_2 \rho_2 + \alpha_3 \rho_3\ ,  & \hspace{1cm} \mbox{[mass-weighted sample]}  \\[0.4cm]
 b_g^{\N} \delta_{\M}
  + \left( \beta_2(M) + \tilde\beta_2(M) \frac{\partial}{\partial\ln\sigma_{\M}} \right) \rho_2 & \\
  \hspace{0.95cm} + \left( \beta_3(M) + \tilde\beta_3(M) \frac{\partial}{\partial\ln\sigma_{\M}} \right) \rho_3\ . & \mbox{[narrow sample]}
\end{array} \right.  \label{eq:gnl_series}
\eeq
To compute power spectra we will need the following cumulants in the $\gnl$ model:
\bea
\xi^{(4)[{\rm tree}]}_{\Phi}
& = & \gnl \big[P_1 P_2 P_3 + \mathrm{23 \, perms.} \big] + \mathcal{O}(\gnl^2)\ , \label{equ:A2g} \\
\xi^{(4)[{\rm loop}]}_{\Phi}
&= & 9 \hskip 1pt\gnl^2  \big[P_{1} P_2 P_{\phi^2}({k}_{13}) + 11 \, {\rm perms.} \big]\ , \label{equ:A3g} \\
\xi^{(6)}_{\Phi}
&= & 36\hskip 1pt \gnl^2 \big[ P_1 P_2 P_3 P_4 P_{125} + 89\, {\rm perms.} \big]\ . \label{equ:A4g}
 \eea
Here, we have defined $P_{ijk} = P_\phi(|\k_i+\k_j+\k_k|)$ and
\beq
P_{\phi^2}(k) \equiv 2 \int_{{\q}} P_\phi(q) P_\phi(|{\k}-{\q}|) \ \sim\ 4 \hskip 1pt \Delta_\phi^2 \ln(kL) P_\phi(k)\ ,
\eeq
where $\Delta^2_{\phi} \equiv (k^3/2\pi^2) P_{\phi}(k)$ and we have regulated the infrared divergence by putting the field in a 
finite box of size $L$.
Note that the power spectra $P_{\delta \rho_2}$ and $P_{\rho_2\rho_3}$ are zero (since there is a $\Phi\rightarrow -\Phi$ symmetry).
The remaining power spectra can be calculated by  substituting eqs.~(\ref{equ:A2g}), (\ref{equ:A3g}) and (\ref{equ:A4g}) into eqs.~(\ref{equ:P1}) and (\ref{equ:P2}). 
In the limit $k \to 0$, this gives
\begin{eqnarray}
P_{\delta \rho_3}(k)
  &=& 3\hskip 1pt \gnl \frac{P_{\m\m}(k)}{\alpha(k)} \kappa_3^{(\fnl=1)} \ ,  \label{eq:gnl1} \\
P_{\rho_2\rho_2}(k)
  &=& \frac{24\hskip 1pt \gnl}{\sigma_{\M}^2}
        \left( \int_{\q} \alpha_{\M}^2(q) P_\phi^2(q) \right) + 36\hskip 1pt \gnl^2 P_{\phi^2}(k)\ , \label{eq:gnl2} \\
P_{\rho_3\rho_3}(k)
  &=& 9\hskip 1pt \gnl^2 \frac{P_{\m\m}(k)}{\alpha^2(k)} \left( \kappa_3^{(\fnl=1)} \right)^2\ .  \label{eq:gnl3}
\end{eqnarray}
Here, $\kappa_3^{(\fnl=1)}$ denotes the dimensionless skewness parameter $\kappa_3 = \langle \delta_{\M}^3(\x) \rangle_{\rm c} / \sigma_{\M}^3$
in the local model with $\fnl=1$.
Note that we use the tree-level cumulant $\xi_\Phi^{(4)[{\rm tree}]}$ when computing $P_{\delta \rho_3}$, but use
both the tree-level cumulant and the one-loop cumulant $\xi_\Phi^{(4)[{\rm loop}]}$ when computing $P_{\rho_2\rho_2}$.  
Although the $\bigoh(\gnl^2)$ one-loop cumulant is generally smaller than the $\bigoh(\gnl)$ tree-level cumulant, 
the one-loop cumulant dominates in the $|\k_1+\k_2| \rightarrow 0$ limit which is
relevant for $P_{\rho_2\rho_2}$.

Putting the above calculations together, we find:\footnote{We have neglected contributions to $P_{\h\h}(k)$ which
approach a constant as $k\rightarrow 0$; such contributions are unobservable in practice since they are
degenerate with other contributions such as second-order halo bias.}
\begin{subequations}
\label{eq:gnl_ps}
\begin{eqnarray}
P_{\m\h}(k) &=& \left( b_g + \gnl \frac{\beta_g}{\alpha(k)} \right) P_{\m\m}(k)\ ,  \\
P_{\h\h}(k) &=& \left( b_g + \gnl \frac{\beta_g}{\alpha(k)} \right)^2 P_{\m\m}(k) + \frac{9}{4} \beta_f^2 \gnl^2 P_{\phi^2}(k) \ , 
\end{eqnarray}
\end{subequations}
where $\beta_f$ was defined in eq.~(\ref{eq:betaf_def}) and we have defined
\beq
\beta_g = \left\{ \begin{array}{lr}
   3 \, \alpha_3(M) \kappa_3^{(\fnl=1)}\ , & \hspace{1cm} \mbox{[mass-weighted sample]} \\
   3 \! \left( \beta_2(M) + \tilde\beta_2(M) \frac{\partial}{\partial\ln\sigma_{\M}} \right) \kappa_3^{(\fnl=1)}\ . & \hspace{1cm} \mbox{[narrow sample]} \\
\end{array} \right.
\eeq
Note that in the narrow mass bin case, there are derivative terms in $\delta_{\h}$ (eq.~(\ref{eq:gnl_series})), 
and their contributions to $P_{\m\h}$ and $P_{\h\h}$ are non-zero (unlike the previously considered $\tnl$ model), because the power spectra
$P_{\rho_m\rho_n}$ in eqs.~(\ref{eq:gnl1})--(\ref{eq:gnl3}) depend on halo mass via the mass-dependent quantity
$\kappa_3^{(\fnl=1)}$.

These expressions for $P_{\m\h}$ and $P_{\h\h}$ agree with previous calculations in the literature 
based on the Edgeworth expansion~\cite{Desjacques:2011mq,Smith:2011ub,stochastic}.
Our series expansion gives some physical intuition as follows.
The non-Gaussian contribution to $P_{\m\h}$ comes from the power spectrum $P_{\delta\rho_3}$,
and can therefore be interpreted as arising from long-wavelength variations in the locally measured
small-scale skewness $\rho_3$.
On large scales, the non-Gaussian fluctuations in $\rho_3$ are 100\% correlated to the density field,
and therefore the associated halo bias is non-stochastic.
The leading contribution to stochastic bias comes from the power spectrum $P_{\rho_2\rho_2}$ and can
be interpreted as long-wavelength variations in small-scale power which are uncorrelated to the
density field.

\section{Proof of the Equivalence}
\label{sec:equivalence}

In the previous section, we showed  that the barrier crossing model can be formulated as a series representation:
\beq
\delta_{\h}(\x) = \left\{ \begin{array}{lr}
  b_g^{\MW} \delta_{\M}({\x}) + \sum_{n \ge 2} \alpha_n(M) \rho_n({\x})\ , & \hspace{0.2cm} \mbox{[mass-weighted sample]} \\
  b_g^{\N} \delta_{\M}({\x}) + \sum_{n \ge 2} \left( \beta_n(M) + \tilde\beta_n(M) \frac{\partial}{\partial\ln\sigma_{\M}} \right) \rho_n({\x})\ .
    & \mbox{[narrow sample]}
\end{array} \right.  \label{eq:lb_series}
\eeq
In this section, we will use this result to prove that barrier crossing is mathematically equivalent to local biasing (\S\ref{sec:LB}) and peak-background split (\S\ref{sec:PBS}).

\subsection{Local Biasing}
\label{sec:LB}

``Local biasing'' refers to any model of halo clustering in which the halo field
is represented as a local function of the dark matter density, e.g.~a power series 
\beq
\delta_{\h}(\x) = b_1 \delta(\x) + b_2 \delta^2(\x) + b_3 \delta^3(\x) + \cdots\ .
\eeq
Several versions of local biasing exist in the literature 
(e.g.~\cite{Fry:1993, Sefusatti:2009, Giannantonio:2009, Baldauf:2011}).
We notice that the series on the right-hand side of (\ref{eq:lb_series}) is a type of local biasing expansion, since the $\rho_n$ fields are local functions of the smoothed density field $\delta_{\M}$.
Therefore, our series representation proves that the barrier crossing model is mathematically equivalent to a specific version of the local biasing formalism.
In this section, we would like to elaborate on the connection between our series representation and the usual way of thinking about
local biasing, and comment on the differences with other versions of the formalism.

First, the density field $\delta_\M$ which appears in the series representation is the non-Gaussian and {\em linearly evolved} density field,
smoothed on the mass scale $M$.  In particular, there is no need to introduce a new smoothing scale which is distinct from the
halo scale, as done in some versions of local biasing.  We do not include non-linear evolution in $\delta_\M$ since the standard
barrier crossing model is based on thresholding the linear density field.

Second, we do not need to introduce explicit dependence of the halo over-density $\delta_{\h}$ on the long-wavelength potential $\Phi_\l$
in a non-Gaussian cosmology.
In some versions of local biasing, $\delta_{\h}$ is expanded in both $\delta_{\l}$ and $\Phi_{\l}$, in order to keep the relation local.
In our version, the $\Phi_{\l}$ dependence happens automatically, since $\delta_{\h}$ depends on higher cumulants $\rho_2, \rho_3, \cdots$,
and these cumulants can be correlated with $\Phi_{\l}$ in a non-Gaussian model.
To see how this happens in detail, consider the $\fnl$ model.
Inspection of the power spectra in eqs.~(\ref{eq:tnl_rho11}) and (\ref{eq:tnl_rho22}) shows (taking $\tnl = (\frac{6}{5} \fnl)^2$) that
$\rho_2$ is 100\% correlated with the field $\Phi_\ell = \alpha_{\M}^{-1}(k) \delta_\M$ as $k \rightarrow 0$.
More precisely, $\rho_2 \rightarrow 4 \fnl \Phi_\ell$ on large scales.
Making this substitution in eq.~(\ref{eq:lb_series}), we get $\delta_{\h} = b_g \delta_{\l} + \fnl \beta_f \Phi_{\l} + \cdots$ 
and recover the usual result.

This example shows that including explicit $\Phi_\ell$ dependence in the local expansion of $\delta_{\h}$ is not necessary 
(in fact, including it our model would ``double-count'' the non-Gaussian clustering), if higher powers of the density field
are included in the expansion.
In the $\fnl$ model, the modulation to the locally measured power $\rho_2$ is directly proportional to $\Phi_{\l}$.
More generally, the expansion should be in all of the non-negligible cumulants $\rho_2, \rho_3, \cdots$.

It is also interesting to consider the $\tnl$ model in the case $\tnl > (\frac{6}{5} \fnl)^2$.  Here, the locally measured small-scale power $\rho_2$
has excess power on large scales which is not 100\% correlated with $\Phi_\ell$, leading to stochastic bias~\cite{stochastic}.  This qualitative behavior is
correctly captured by a local biasing model of the form $\delta_{\h} = b_g \delta_\l + \alpha_2 \rho_2$, but not by a local biasing model of the
form $\delta_{\h} = b_g \delta_\l + b_2 \Phi_\l$.

In the narrow mass bin case, our series expansion includes derivative terms of the form $\partial \rho_n / \partial\ln\sigma_{\M}$.
To our knowledge, derivative terms have not been been proposed in any version of local biasing which has appeared in the literature.
In the barrier crossing model, derivative terms appear naturally for a narrowly selected halo sample, since this case is obtained from the
mass-weighted case (which does not contain derivative terms) by differentiating with respect to halo mass.

Finally, even in the mass-weighted case, there is a difference between the Hermite polynomial expansion
\beq
\delta_{\h}({\x}) = b_g^{\MW} \delta_{\M}({\x}) + \sum_{n\ge 2} \alpha_n(M)\hskip 1pt  H_n\left( \frac{\delta_{\M}({\x})}{\sigma_{\M}} \right)  \label{eq:hermite_series}
\eeq
and a power series expansion of the form
\beq
\delta_\h(\x) = b_1 \delta_\M(\x) + b_2 \delta_{\M}^2(\x) + b_3 \delta_{\M}^3(\x) + \cdots  \ . \label{eq:power_series}
\eeq
At first sight, the two may appear equivalent: if both series are truncated at the same order $N$, then we can rearrange coefficients to transform
either series into the other (since both just parametrize an arbitrary degree-$N$ polynomial).
However, when we write the power series expansion~(\ref{eq:power_series}), we are assuming that the values of the low-order coefficients $b_1, b_2, \cdots$
are independent of the order $N$ at which the series is truncated.
This means for example that in a Gaussian cosmology, the matter-halo power spectrum 
$P_{\m\h}(k) = (b_1 + 3 \hskip 1pt \sigma_{\M}^2 b_3 + 15\hskip 1pt \sigma_{\M}^4 b_5 + \cdots) P_{\m\m}(k)$
depends on where the series is truncated.
In contrast, the Hermite expansion~(\ref{eq:hermite_series}) is more stable: $P_{\m\h}(k)$ is always equal to $b_g^{\MW} P_{\m\m}(k)$,
regardless of how many terms are retained in the series.
Note that the barrier crossing model has a convergent Hermite polynomial expansion~(\ref{eq:herm2}), but cannot be sensibly expanded as a power series in $\delta_\M$,
since the Heaviside step function $\Theta(\delta_{\M}/\sigma_{\M} - \delta_c)$ is not an analytic function of $\delta_{\M}$.

In summary, the barrier crossing model is mathematically equivalent to a specific version of the local biasing formalism in which the
following choices have been made: we linearly evolve the density field and smooth it at mass scale $M$; we include higher cumulants
$\rho_2, \rho_3, \cdots$ in the density field, but not additional fields such as the potential $\Phi_\l$; derivative terms appear in
the narrow mass bin case; and we use a Hermite polynomial expansion
in $\delta_\M/\sigma_\M$ rather than the power series expansion.
Other variants of the local biasing formalism exist in the literature, and 
we are not claiming that our choices are optimal (in the sense of producing best agreement with simulations);
the purpose of this section was simply to point out which set of choices is equivalent to the barrier crossing model.

\subsection{Peak-Background Split}
\label{sec:PBS}

The ``peak-background split" is a formalism for modeling halo clustering on large scales, in which one relates large-scale modes of the
halo density field $\delta_\h$ to large-scale modes of fields whose power spectra can be calculated directly.
For example, the PBS formalism was applied to an $\fnl$ cosmology in~\cite{Slosar:2008hx}.
On large scales, $k \ll R_{\M}^{-1}$, one can argue that the halo density is related to the linear density field $\delta$
and the Newtonian potential $\Phi$ by
\beq
\delta_\h(\k) = b_g \delta(\k) + \fnl \beta_f \Phi(\k)\ ,
\eeq
where $b_g$ is the usual Gaussian bias, and $\beta_f = 2 \hskip 1pt \partial\ln n_\h / \partial\ln\sigma_8$.
Using this expression, it is easy to show that the large-scale bias is given by $b(k) = b_g + \fnl \beta_f / \alpha(k)$,
and is non-stochastic.  For additional examples of the PBS formalism applied to non-Gaussian models, 
see~\cite{Tseliakhovich:2010kf,Smith:2011ub,stochastic}.
In this section, we will show how the PBS formalism generalizes to an arbitrary non-Gaussian model,
and give a simple proof that this generalization is equivalent to the barrier crossing model.
We will work out in detail the case of a mass-weighted halo sample; the narrow mass bin case follows by differentiating
with respect to $M$.

\vskip 4pt
There is one technical point that we would like to make explicit.
We want to generalize the peak-background split formalism so that it applies to an arbitrary non-Gaussian model,
parametrized by the $N$-point correlation functions of the initial Newtonian potential $\Phi$.
As an example, consider the $\tnl$ model from \S\ref{ssec:tnl}, with constituent fields $\phi, \psi$.
The PBS analysis of this model has been worked out in~\cite{Tseliakhovich:2010kf,stochastic} and requires keeping 
track of the long-wavelength parts $\phi_\ell, \psi_\ell$ of both fields, in order to correctly predict
non-Gaussian stochastic bias on large scales.
(Intuitively, multiple fields are needed because we need to keep track of long-wavelength density fluctuations and long-wavelength
variations in the locally measured small-scale power, and the two are not 100\% correlated in the $\tnl$ model.)
This raises a conceptual puzzle: how would we get stochastic bias if we were just given correlation functions of the single field $\Phi$, 
rather than a description of the $\tnl$ model involving multiple constituent fields?
As we will now see, we must extend the PBS formalism by introducing additional fields
which correspond to the locally measured small-scale power, small-scale skewness, kurtosis, etc.
These fields are precisely the quantities $\rho_2, \rho_3, \cdots$ which appeared earlier in our
series expansion in~\S\ref{sec:series_representation}.  This will allow us to connect the PBS formalism with the barrier crossing model
(and in fact prove that the two are mathematically equivalent).

\vskip 4pt
Consider a large subvolume of the universe containing many halos, but over which the long mode is reasonably constant,
and let $( \cdot )_\ell$ denote a spatial average over the subvolume.
Let us assume that the halo number density $(n_\h)_\ell$ in the subvolume is a function of the one-point PDF of the underlying 
dark matter field $\delta_\M$ (when linearly evolved and smoothed on the halo scale). 
For weakly non-Gaussian fields, the one-point PDF in each subvolume can be characterized completely by its mean $(\delta_\M)_\ell$, variance
$(\sigma^2_\M)_\ell$, and higher cumulants $(\kappa_n)_\ell = (\langle \delta^n_\M \rangle_{\rm c}/\sigma^n_\M)_\ell$ for $n \ge 3$. Therefore we can write $(n_\h)_\ell \equiv \bar n_\h ( (\delta_\M)_\ell, (\sigma^2_\M)_\ell, \{(\kappa_n)_\ell\})$.
Taylor expanding to \textit{first order} in these parameters, we get
\beq
(n_\h)_\ell = \bar n_\h \left( 1 
  + \frac{\partial\ln n_\h}{\partial (\delta_\M)_\ell}\, (\delta_\M)_\ell
  + \frac{\partial\ln n_\h}{\partial (\sigma^2_\M)_\ell} \big((\sigma^2_\M)_\ell - \sigma_{\M}^2 \big)
  + \sum_{n=3}^\infty \frac{\partial\ln n_\h}{\partial (\kappa_n)_\ell} \, (\kappa_n)_\ell
\right) \ . \label{eq:PBS_exp0}
\eeq
Here, we have used the notation $(\sigma^2_\M)_\ell$ to denote the variance of $\delta_{\M}$ restricted to the subvolume, and $\sigma_{\M}^2$
to denote the global variance.
To make contact with our previous notation, note that 
$\big((\sigma^2_\M)_\ell - \sigma_{\M}^2\big) = \sigma_{\M}^2 (\rho_2)_\ell$ and $(\kappa_n)_\ell = (\rho_n)_\ell$.\footnote{The 
identity $(\rho_n)_\ell = (\kappa_n)_\ell$ holds for $n\le 5$, but has non-linear corrections for $n\ge 6$.
For example, $(\rho_6)_\ell = (\kappa_6)_\ell + 10 (\kappa_3)_\ell^2$.
We have neglected these non-linear corrections since eq.~(\ref{eq:PBS_exp0}) is an expansion to first order anyway.}
Making these substitutions in eq.~(\ref{eq:PBS_exp0}), we get
\beq
(\delta_\h)_\ell =
  \frac{\partial\ln n_\h}{\partial (\delta_\M)_\ell} (\delta_\M)_\ell
  + \sigma_{\M}^2 \frac{\partial\ln n_\h}{\partial (\sigma_{\M}^2)_\ell} \, (\rho_2)_\ell
  + \sum_{n=3}^\infty \frac{\partial\ln n_\h}{\partial (\kappa_n)_\ell} \, (\rho_n)_\ell\ .
\eeq
Since this equation applies when taking the subvolume average $(\cdot)_\ell$ over any large subvolume, it also
applies to any large-scale Fourier mode:
\beq
\delta_\h(\k) \ \xrightarrow{k\to 0} \
  \frac{\partial\ln n_\h}{\partial \delta_\M} \delta_\M(\k)
  + \sigma_{\M}^2 \frac{\partial\ln n_\h}{\partial \sigma^2_\M}\, \rho_2(\k)
  + \sum_{n=3}^\infty \frac{\partial\ln n_\h}{\partial \kappa_n}\, \rho_n(\k)\ .  \label{eq:pbs_ser1}
\eeq
Let us compare this expression with our series representation of $\delta_{\h}$ in the barrier crossing model:
\beq
\delta_{\h}(\k) = b_g^{\MW} \delta_{\M}(\k) + \sum_{n \ge 2} \alpha_n(M) \rho_n(\k) \ . \label{eq:pbs_ser2}
\eeq
The form of the two series representations is the same, but the coefficients appear to be different.
In the barrier crossing model, we have the following explicit formula
for the coefficient $\alpha_n(M)$ of the $n$-th term in the series:
\beq
\alpha_n(M) = \sqrt{ \frac{2}{\pi} }\, \frac{e^{-\nu_c^2/2}}{{\rm erfc}(\frac{1}{\sqrt{2}} \nu_c)}\, \frac{H_{n-1}(\nu_c)}{n!} \ , \label{eq:alpha_barrier}
\eeq
whereas in the PBS derivation, $\alpha_n$ is given by a suitable derivative of the halo mass function:
\beq
\alpha_2 = \sigma_{\M}^2 \frac{\partial\ln n_\h}{\partial \sigma^2_\M} \hspace{0.4cm} {\rm and\ } \hspace{0.4cm}
\alpha_n = \frac{\partial\ln n_\h}{\partial \kappa_n} \hspace{0.3cm} {\rm for\ } n\ge 3\ .  \label{eq:alpha_pbs}
\eeq
If we assume a Press-Schechter mass function, then one can evaluate the mass function derivatives in the above equation using the 
machinery from~\cite{LoVerde:2007ri}.
The result agrees precisely with the explicit formula~(\ref{eq:alpha_barrier}).
Therefore, the barrier crossing model and the generalized PBS formalism with fields $\rho_2,\rho_3,\cdots$
are formally equivalent, but only under the assumption of a Press-Schechter mass function 
(note that this assumption is ``built in'' to the barrier crossing model).

If we relax the assumption of a Press-Schechter mass function, then the barrier crossing model and
the generalized PBS formalism can both be written as series expansions with the same general form,
but make different predictions for the coefficients $\alpha_n(M)$.
One can ask which prediction agrees better with $N$-body simulations.
In~\cite{Smith:2011ub}, the two predictions for $\alpha_3$ were compared with simulations in the context of the $\gnl$ model.
It was found that the PBS prediction~(\ref{eq:alpha_pbs}) is exact (within the $\approx 1$\% statistical error of the simulations)
if both the bias and the mass function derivative $(\partial\ln n_\h/\partial\kappa_3)$ are evaluated numerically from the 
simulations.
The barrier crossing prediction~(\ref{eq:alpha_barrier}) is an approximation: 
although it is based on an exact calculation within the barrier crossing model, this model is an
approximation to the true dynamics of an $N$-body simulation.
The approximation works reasonably well for large halo mass but breaks down for low masses,
motivating the use of fitting functions for practical data analysis.
It is natural to conjecture that the same qualitative statements will be true for the $\alpha_n$ coefficients with $n > 3$,
but we have not attempted to verify this with simulations. 
(Note that no fitting function is necessary for $\alpha_2$, since the relation $\beta_f \approx 2 \delta_c b_g$ holds to
$\approx 10$\% accuracy in $N$-body simulations.)

In summary, the barrier crossing model is mathematically equivalent to the PBS formalism, appropriately generalized
to an arbitrary non-Gaussian cosmology by introducing additional fields $\rho_2, \rho_3, \cdots$, plus the additional
assumption of a Press-Schechter mass function.
The barrier crossing model is analytically tractable 
(e.g.~one can derive closed-form expressions for the coefficients $\alpha_n(M)$ and $\beta_n(M)$), 
and usually a reasonable approximation, making it very useful for analytic studies or forecasts.
However, for data analysis, it may be necessary to go beyond the Press-Schechter approximation by
replacing the closed-form expressions for coefficients such as $\alpha_n(M)$ with their PBS counterparts measured from simulations.



\section{Conclusions}
\label{sec:discussion}


In this paper, we have proven the mathematical equivalence of barrier crossing, peak-background split and local biasing.
We first introduced a Hermite polynomial expansion of the halo density contrast $\delta_{\h}$ in the barrier crossing model: eqs.~(\ref{eq:series_mw}) and~(\ref{eq:series_narrow}).
We showed that this allows a computationally efficient way to calculate the clustering power spectra $P_{\m\h}$ and $P_{\h\h}$.  
Moreover, the series expansion makes the formal equivalence of the various halo modeling formalisms very transparent. First, it automatically takes the form of a local biasing model, in which the non-Gaussian and linearly evolved density contrast is expanded in Hermite polynomials. 
Second, it provides a very natural connection between barrier crossing and peak-background split.
To make this relationship manifest, we generalized the PBS formalism so that it can be applied to the most general set of non-Gaussian initial conditions, parametrized by the $N$-point functions of the primordial potential. 
This extension of PBS involves additional fields which correspond to the locally measured small-scale power, small-scale skewness, kurtosis, etc.
Mapping those fields to fields in the Hermite polynomial expansion of the barrier crossing model, we proved the mathematical equivalence between PBS and BC.
Finally, although, in this paper, we have concentrated on computing power spectra, our series expansion should also be useful for analyzing the effects of primordial non-Gaussianity on other clustering statistics, such as the halo bispectrum~\cite{Baldauf:2011}.

\newpage
\acknowledgments
We thank Marilena LoVerde, Marcel Schmittfull, David Spergel and Matias Zaldarriaga for helpful discussions.
S.F.~acknowledges support from a fellowship at the Department of Astrophysical Sciences of Princeton University.
K.M.S.~was supported by a Lyman Spitzer fellowship in the Department of Astrophysical Sciences at Princeton University. Research at Perimeter Institute is supported by the Government of Canada
through Industry Canada and by the Province of Ontario through the Ministry of Research \& Innovation.
The research of D.G.~is supported by the DOE under grant number DE-FG02-90ER40542 and the Martin A.~and Helen Chooljian Membership at the Institute for Advanced Study.
D.B.~gratefully acknowledges support from a Starting Grant of the European Research Council (ERC STG grant 279617). 

\appendix
\section{Hermite Polynomials}
\label{sec:hermite}

In this paper, we have used the {\it probabilists'} definition of Hermite Polynomials
\beq
H_n(\nu) = (-1)^n e^{\nu^2/2} \frac{d^n}{d \nu^n} e^{-\nu^2/2}\ ,
\eeq
satisfying the recursion relation
\beq
H_{n+1}(\nu) = \nu H_{n}(\nu) - H'_{n}(\nu)
\eeq 
and the orthogonality condition
\beq
\int_{-\infty}^\infty \d \nu\, \frac{1}{\sqrt{2\pi}}e^{-\nu^2/2}\, H_m(\nu) H_n(\nu) = m! \hskip 1pt \delta_{mn}\ .
\eeq
For reference, we list some of the low-order Hermite polynomials
\begin{align}
H_0(\nu) &= 1\ ,\\
H_1(\nu) &= \nu\ , \\
H_2(\nu) &= \nu^2-1\ , \\
H_3(\nu) &= \nu^3-3\nu\ , \\
H_4(\nu) &= \nu^4-6\nu^2+3\ . 
\end{align}
We have made use of the following integral
\beq
 \frac{1}{n!} \int_{\nu_c}^\infty \d\nu \, \frac{1}{\sqrt{2\pi}}e^{-\nu^2/2}\, H_n(\nu) 
  = \left\{ \begin{array}{ll} 
        \tfrac{1}{2}{\rm erfc}(\tfrac{1}{\sqrt{2}} \nu_c) & \quad n=0 \\ 
        \tfrac{1}{n!} \frac{1}{\sqrt{2\pi}} e^{-\nu_c^2/2}\, H_{n-1}(\nu_c)& \quad n\ge 1
    \end{array} \right. \ . 
    \eeq

\newpage
 \begingroup\raggedright\endgroup

\end{document}